\documentclass[journal=nalefd,manuscript=letter]{achemso}

\usepackage{chemformula} 
\usepackage{soul, color}
\usepackage{wasysym}
\usepackage{natbib}
\usepackage{mathtools}
\usepackage{subfigure}  
\usepackage{amsmath}
\usepackage{caption}
\usepackage{soul,color}

\author{Julia Frank}
\email{umfran77@myumanitoba.ca}
\affiliation{Department of Physics \& Astronomy, University of Manitoba, Winnipeg, Canada}

\author{Johan van Lierop}

\affiliation{Department of Physics \& Astronomy, University of Manitoba, Winnipeg, Canada}

\author{Robert L. Stamps}
\email{Robert.Stamps@umanitoba.ca}
\affiliation{Department of Physics \& Astronomy, University of Manitoba, Winnipeg, Canada}

\title{Monopole current control in artificial spin ice via localized fields} 


\begin{document}

\begin{abstract}

Artificial spin ice systems are metamaterials composed of interacting nanomagnets arranged on a lattice, exhibiting geometrical frustration and emergent phenomena such as monopole excitations. We explore magnetization dynamics and monopole current control in square artificial spin ice with added vertical control elements. Using Monte Carlo simulations, we examine how localized magnetic fields from these elements influence vertex configurations and domain propagation, enabling directional and polarity control of monopole currents. The control elements suppress monopole nucleation along one edge, steering monopole flow across the lattice -- sometimes even against the applied field direction. These elements also reshape the system’s energy landscape, producing tailored hysteresis and guided state transitions. Our results offer a strategy for manipulating collective behaviours in artificial spin ice using localized fields. This has implications for magnetic memory, physical reservoir computing, enabling reconfigurable magnetic logic and spin-based information processing, and device architectures requiring directional magnetic charge transport.

\end{abstract}

\newpage

Artificial spin ices are metamaterials made with nanomagnets arranged in a lattice structure\cite{wang_artificial_2006, moller_artificial_2006}.  The nanomagnets are single-domain \emph{macro}spins where each behaves much like a compass needle and is typically composed of a magnetically `soft', low magnetocrystalline anisotropy material such as Ni$_{80}$Fe$_{20}$, which permits straightforward reversal of nanomagnet magnetization with an applied magnetic field.  The nanomagnets interact through dipolar fields collectively and are arranged so that geometrical frustration is created\cite{nisoli_shiffer_imaging_2013, Cumings_2014, saccone_cairoLat_2019}.  Thermodynamic systems can be made where the energy barriers to nanomagnet macrospin reversals can occur spontaneously 
at a set temperature, with reversal probabilities determined by the dipolar interactions.  Overall, artificial spin ice systems provide a rich playground for the investigation of frustration, topological structures like magnetic monopoles and their dynamics, and novel topological phases and phase transitions\cite{nisoli_shiffer_imaging_2013, perrin_extensive_2016, sklenar_field-induced_2019, anghinolfi_thermodynamic_2015, skjaervo_advances_2020}.

Lattices of macrospins can be decomposed into a tiling of adjacent macrospins. Each tile is commonly referred to as a \emph{vertex}. This vertex-based perspective is helpful in understanding the collective behaviour\cite{leo_collective_2018} of macrospins and emergent properties arising from interactions at the lattice vertexes. Thermal fluctuations affect vertex frustrations\cite{harris_1997, nisoli_2010}, monopole currents, and macrospin reversals with formation of vertex domains, and nucleation and propagation of these domains within artificial spin ice\cite{sendetskyi_phases_2019}. Vertex domains refer to regions composed of neighbouring vertexes of the same type, defined by their configurations\cite{wang_artificial_2006, Levis2012TwodimensionalSI, saccone_cairoLat_2019, skjaervo_advances_2020}. 

\begin{figure*}[t!]

	\begin{subfigure}{\textbf{(a)}}
        %\centering
        \includegraphics[width=0.44\linewidth]{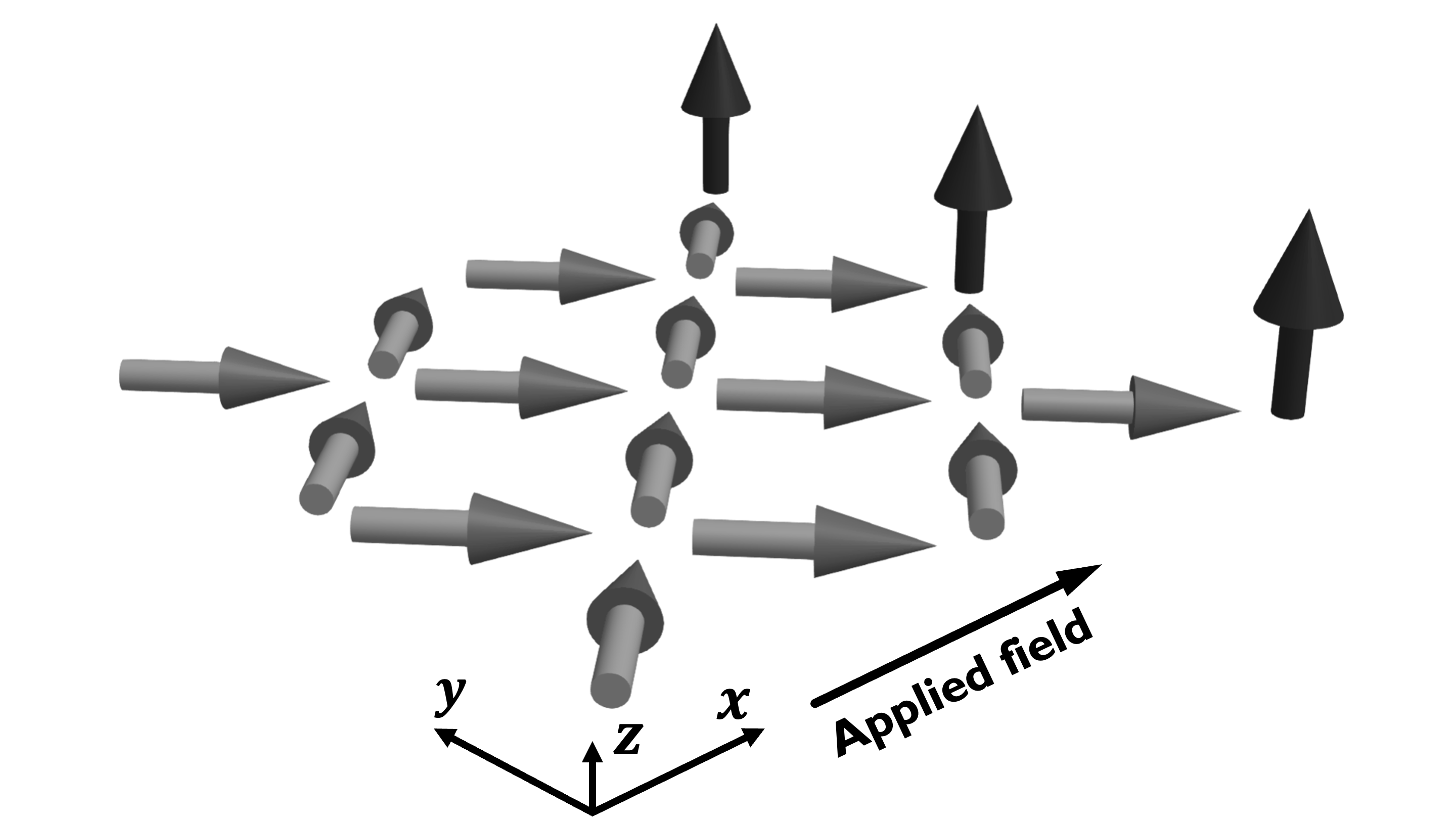}
	\end{subfigure}
	\hfill
	\begin{subfigure}{\textbf{(b)}}
	%\centering
        \includegraphics[width=0.44\linewidth]{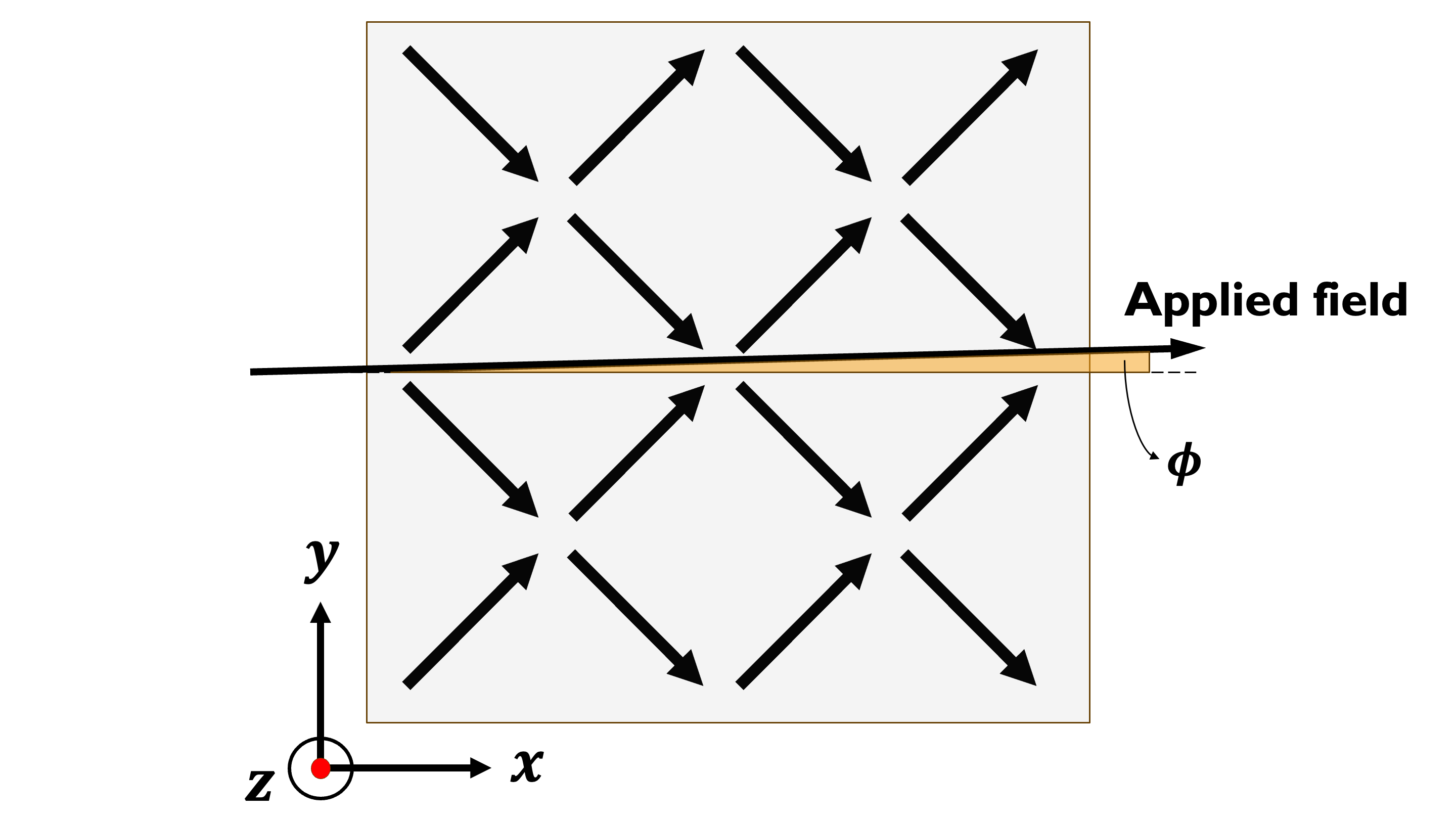}
	\end{subfigure}
	\vfill
	\begin{subfigure}{\textbf{(c)}}
        %\centering
        \includegraphics[width=0.33\linewidth]{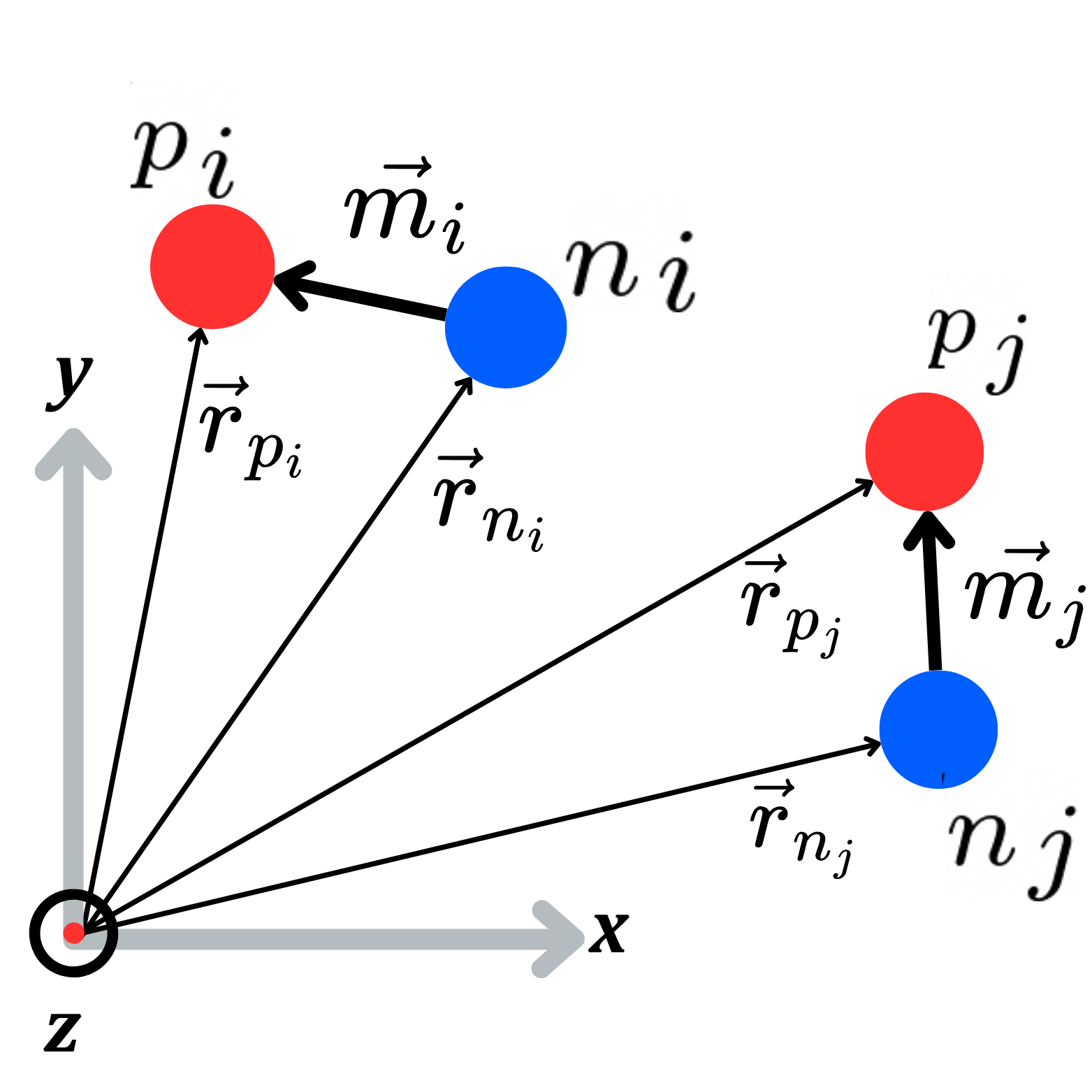}
        \end{subfigure}
        \hfill
        \begin{subfigure}{\textbf{(d)}}
        %\centering
        \includegraphics[width=0.55\linewidth]{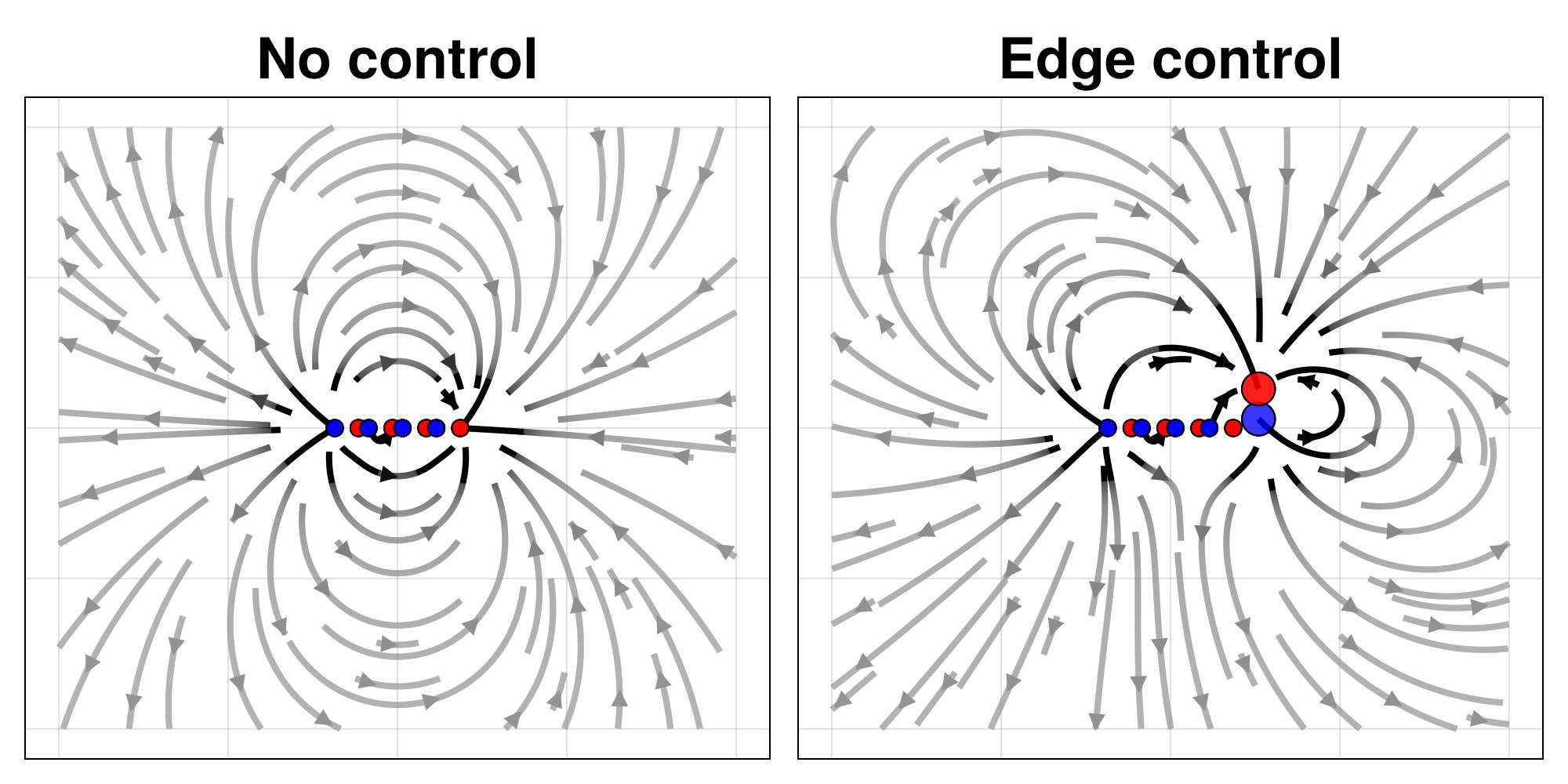}
        \end{subfigure}
        
\caption{\textbf{(a)} 3D representation of the simulated square artificial spin ice system with control macrospins. The figure shows the arrangement of in-plane lattice macrospins (gray arrows), with the vertical control macrospins (black arrows) extending out of the lattice plane and perpendicular to it.  \textbf{(b)} Top view of the system, emphasizing the small symmetry-breaking applied field angle $\phi$=0.1$^{\circ}$ with respect to the $x$-axis. \textbf{(c)}  A pair of interacting artificial spin ice elements represented as dumbbells with magnetic north pole (red) and south pole (blue) of pole strength $q_{p_i}$ and $q_{n_i}$, $q_{p_j}$ and $q_{n_j}$, where subscripts $p$ and $n$ stand for positive and negative charge polarity. The vectors $\vec{r}_{p_i}$ and $\vec{r}_{n_i}$ represent the positions of the positive and negative poles associated with element $i$, respectively, while $\vec{r}_{p_j}$ and $\vec{r}_{n_j}$ represent those for element $j$. The vectors $\vec{m}_i$ and $\vec{m}_j$ denote the magnetic moment vectors (macrospins) of elements $i$ and $j$. \textbf{(d)} The stray field distribution around the system, as seen in the $xz$-plane, due to the “free” uncompensated magnetic poles around the edges. The system of lattice macrospins corresponding to the square spin ice elements is magnetized in the positive $x$-direction.
\label{fig:experimental_setup}}
 
\end{figure*}

We propose introducing control over local magnetic fields from nanomagnets whose macrospins point perpendicular to the lattice macrospins and lie along a lattice edge, as shown in \textbf{figure~\ref{fig:experimental_setup}a}. We explore this approach with a modelled square artificial spin ice structure represented by horizontal arrows in \textbf{figures~\ref{fig:experimental_setup}a} and \textbf{b}. Vertical edge `control' nanomagnets in the figure are represented by the vertical macrospins, and an external applied field in the plane of the lattice is used to control response at finite low temperature and identify experimentally accessible signatures of controlled vertex dynamics. The dimensions and material of the controls are assumed to be the same as those of the lattice nanomagnets.

The general behaviour of the lattice with and without control macrospins was studied using a single spin flip Monte Carlo algorithm\cite{porro_magnetization_2019} (see \textbf{Notes}) on a 16 macrospin square ice that was at a temperature $T$=0.3 (relative energy units\cite{goncalves_tuning_2019}) well below the `ordering' temperature $T_C$=2.3\cite{poppy_MScThesis_2022}.

The overall artificial spin ice magnetism is from the combined effects of temperature and applied field on macrospin reversal subject to their dipolar interactions.  There will be a change in energy from each macrospin's magnetization orientation reversal, which results in the flipping of a macrospins' north and south poles. We describe the dipolar interactions between macrospin pairs using a dumbbell model\cite{castelnovo_magnetic_2008, Castelnovo_2012, Nascimento.2024, may_saccone_magnetic_2021}, that accounts for the finite size of the artificial spin ice elements through the separation of their magnetic poles. The associated energy change for each macrospin flip is calculated for each Monte Carlo sampling step, and is described by
\begin{equation}
E = \frac{\mu_{0}}{4\pi}\sum_{{i,j}}(E_{p_{i}p_{j}} + E_{n_{i}n_{j}} + E_{p_{i}n_{j}} + E_{p_{j}n_{i}}) - \mu_{0}\sum_{i}\vec{m}_{i}\cdot\vec{H}
\label{eqn:energy}
\end{equation}
with interactions between macrospin pairs $i$ and $j$ in the lattice (\textbf{figure~\ref{fig:experimental_setup}c}) given by
\begin{equation}
E_{p_i p_j} = \frac{q_{p_i}q_{p_j}}{|\vec{r}_{p_i} - \vec{r}_{p_j}|}, \;\; E_{n_i n_j} = \frac{q_{n_i}q_{n_j}}{|\vec{r}_{n_i} - \vec{r}_{n_j}|} \;\; E_{p_i n_j} = \frac{q_{p_i}q_{n_j}}{|\vec{r}_{p_i} - \vec{r}_{n_j}|} \;\;  E_{p_j n_i} = \frac{q_{p_j}q_{n_i}}{|\vec{r}_{p_j} - \vec{r}_{n_i}|} \;\;.
\label{eqn:dumbbell}
\end{equation}
The first term in equation~\ref{eqn:energy} is the \emph{dumbbell energy} due to the relative arrangement of the macrospins in the lattice, and the second term is energy of the system in the external applied field. 
To avoid artifacts arising from degeneracy due to $\mu_0 \vec H$ being aligned perfectly with macrospins in the lattice plane, we introduce a symmetry-breaking angle $\phi=0.1^{\circ}$  (Fig.~\ref{fig:experimental_setup}b).

\begin{figure*}[t!]

	\begin{subfigure}{\textbf{(a)}}
        %\centering
        \includegraphics[width=0.95\linewidth]{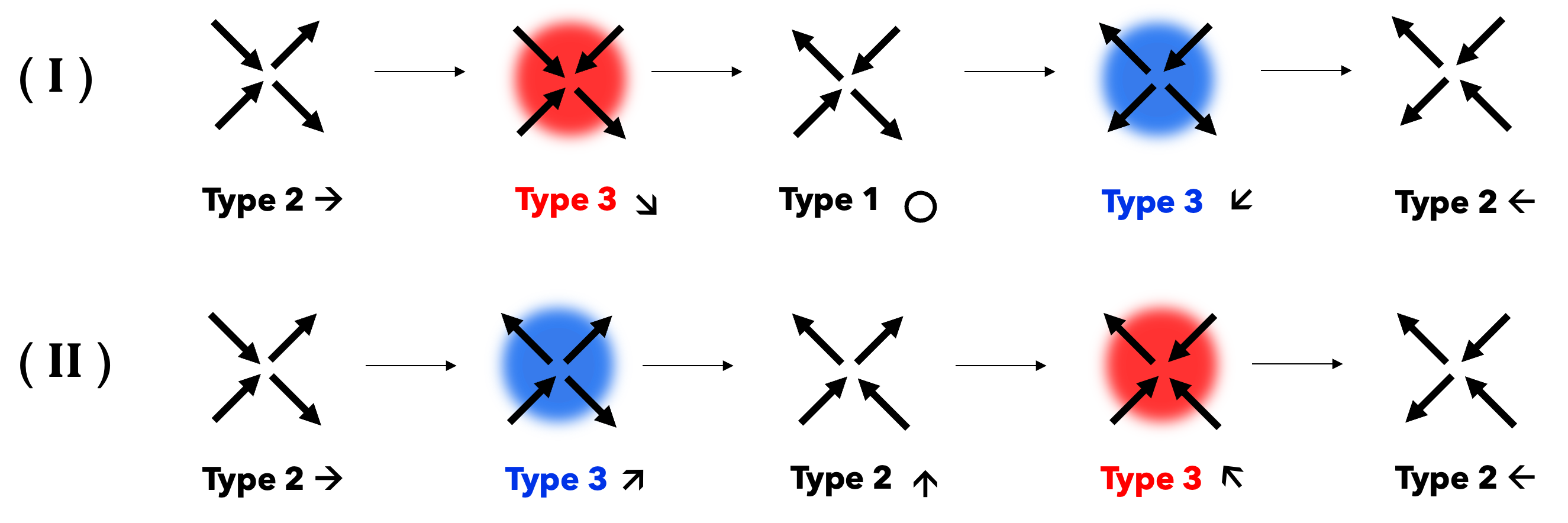}
	\end{subfigure}
	
	\vspace*{12pt}
	\begin{subfigure}{\textbf{(b)}}
        %\centering
        \includegraphics[width=0.95\linewidth]{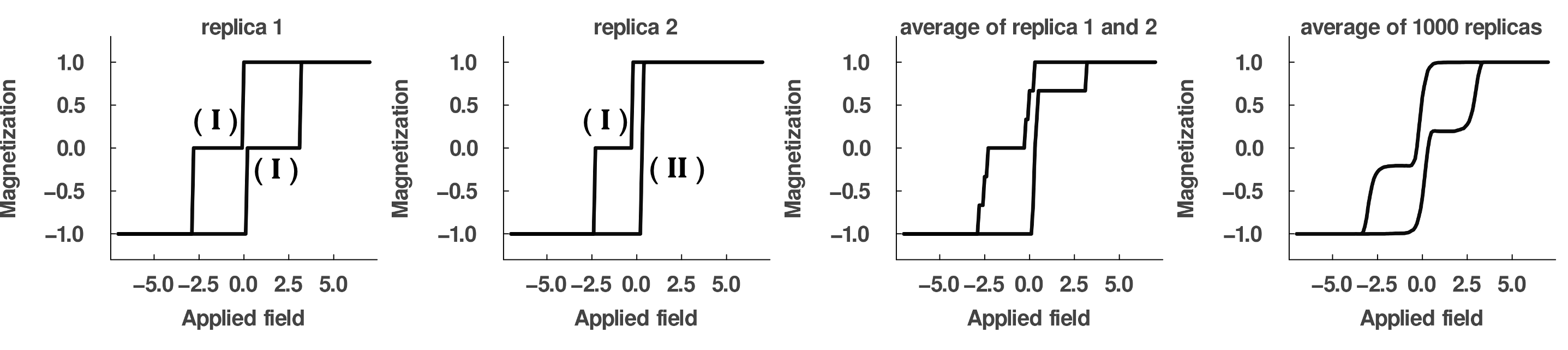}
        \end{subfigure}
        
\caption{ Four-spin ice. \textbf{(a)} Two Type 2 vertex reversal processes are shown as the applied field changes from positive saturation to negative saturation (left-to-right sequence). The labels below each four-spin vertex configuration indicate the net vertex magnetization and its direction. The symbol ($\bigcirc$) corresponds to zero net magnetization, which is characteristic of the Type 1 vertex.  For other vertex types, the labels (e.g., `→', `←') denote the direction of the net vertex magnetization vector.\textbf{(b)} Field-dependence of the overall vertex magnetization is shown for a Type 2 vertex undergoing reversal process I on both loop branches (replica 1), a combination of reversal processes I and II (replica 2), the result of replicas 1 and 2 together, and the overall magnetization of 1000 replicas (field-cycles).\label{fig:T2_reversal}}

\end{figure*}

We first consider the field-dependent behaviour of the four nanomagnets that make up a vertex in square ice as they are the building blocks that are tuned via control macrospins and where monopoles emerge. The four macrospins of a vertex in square ice exhibit $2^4$=16 different configurations (see \textbf{Supporting Information Figure~S1}) that are described according to four vertex types\cite{Levis2012TwodimensionalSI} each with a multiplicity factor. The energies of these vertex macrospin configurations order as $E_{Type 1} < E_{Type 2} < E_{Type 3} < E_{Type 4}$\cite{BUDRIKIS2014109}, with Type 1 the ground-state configuration.  Excitations that take the system out of the ground state are from Type 2 and 3 vertexes, while Type 4 are rare and present only near the ordering temperature. Type 3 vertex configurations are of special interest as they are nonzero magnetization and are said to carry the monopole charge\cite{LEON201659, duarte_magnetic_2024}.

The field-dependent magnetization of the vertex experiences hysteresis, stepping through a Type 2 $\rightarrow$ Type 3 $\rightarrow$ \textbf{Type 1} $\rightarrow$ Type 3 $\rightarrow$ Type 2 vertex macrospin path  (\textbf{figure~\ref{fig:T2_reversal}a} process I) or, equally likely, undergoing a Type 2 $\rightarrow$ Type 3 $\rightarrow$ \textbf{Type 2} $\rightarrow$ Type 3 $\rightarrow$ Type 2 vertex macrospin path (Fig.~\ref{fig:T2_reversal}a process II).  The resulting field-dependent magnetization can be visualized as hysteresis loop-like behaviour corresponding to the two processes. The stochastic average over field-dependent magnetization values is presented in \textbf{figure~\ref{fig:T2_reversal}b} as an average taken over 1000 replicas, where each replica is an identical system of macrospins, evolving independently of the other replicas. The apparent field-shift of the magnetization is an artifact due to excited corner vertex states that generate stray fields, which are not present for ground state Type 1 vertexes due to flux closure. This four-spin analysis forms the basis for understanding collective behaviour in larger lattices.

\begin{figure*}[t!]

        \begin{subfigure}{\textbf{(a)}}
        \includegraphics[width=0.95\linewidth]{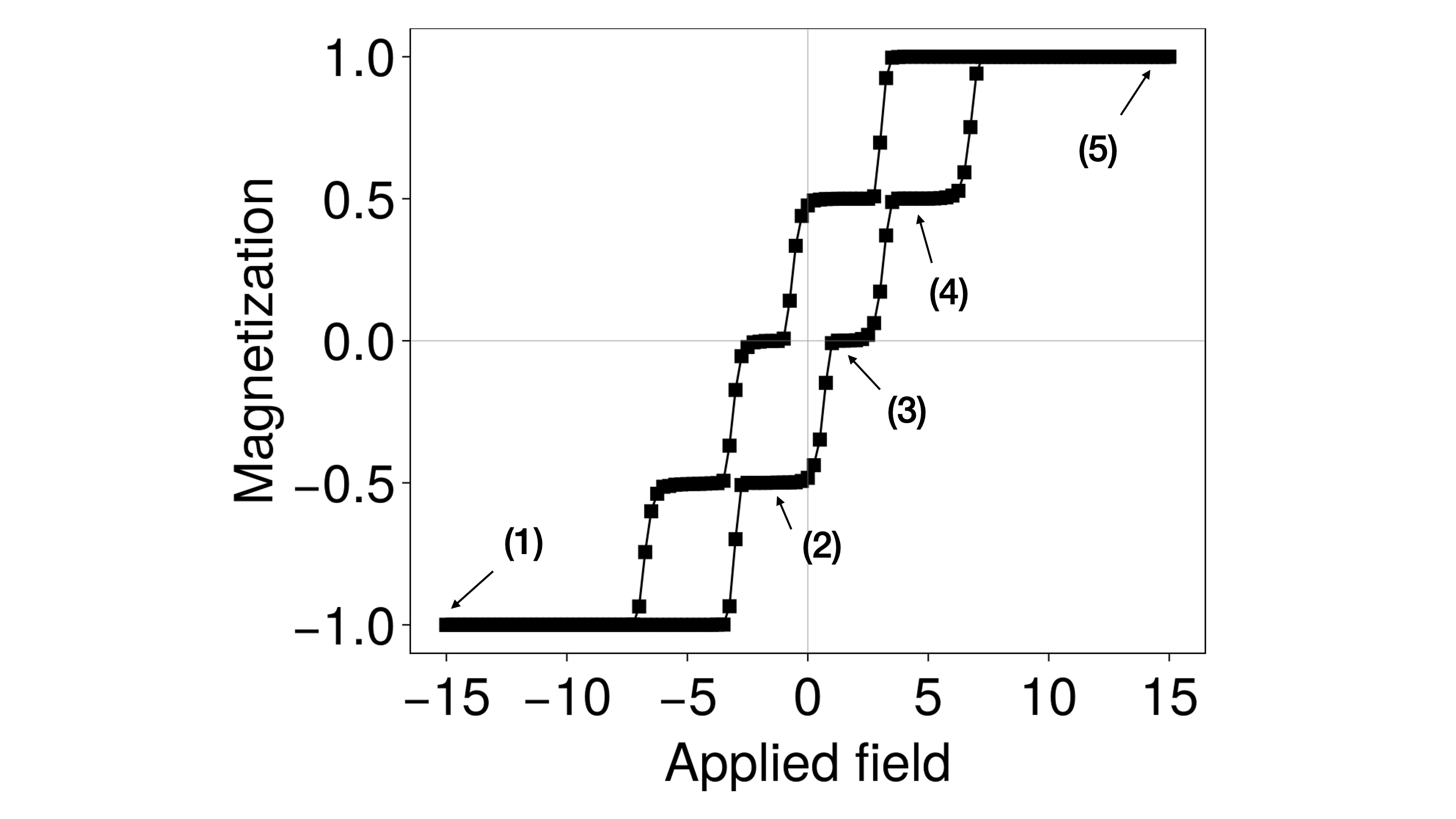}
        \end{subfigure}
        
        \begin{subfigure}{\textbf{(b)}}
        \includegraphics[width=0.95\linewidth]{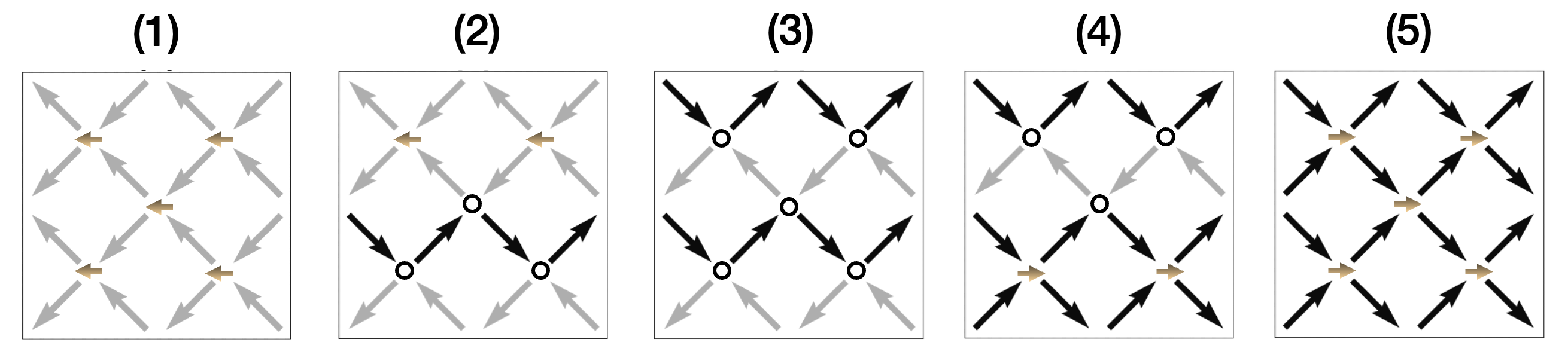}
        \end{subfigure}

\caption{16-spin ice. \textbf{(a)} Average field-dependence of the magnetization for 1000 replicas with representative macrospin and vertex configurations over the field regimes 1--5 are presented in \textbf{(b)} and shown for one of the replicas. The magnetization of the square ice lattice is in plane and parallel to the field. The magnetization calculated as $M = \sum_{i}(\vec{m}_{i}\cdot\vec{H}) / \frac{1}{2} N [\ \cos{(45^{\circ}+\phi)}+\cos{(45^{\circ}-\phi} ) ]\ $, with $N$ the number of macrospins.  Left- or right-oriented brown arrow markers signify a Type 2 vertex ($\leftarrow$ and $\rightarrow$), and the black circle markers signify a Type 1 vertex ($\bigcirc$) to show \emph{vertex domains} alongside the macrospin domains.\label{fig:No_control_SASI}}

\end{figure*}

The step structure observed for the four-spin ice persists when the size is increased to a 16-spin ice. Starting with a large applied field forces the system to a Type 2 tiling of vertexes.  As the field is reduced, macrospin reversals occur via thermal activation. This introduces several new macrospin configurations and a discrete step-like magnetization progression over a range of applied fields due to vertex domains forming with identical magnetizations. An average over 1000 replicas is shown in \textbf{figure~\ref{fig:No_control_SASI}a}. Individual replicas from the average are shown in \textbf{SI2}. Vertex reversal processes I and II occur, with the Type 3 vertex being the field-nucleation `trigger' site in the lattice.  Macrospin propagation along the applied field direction occurs at an edge macrospin as an field-driven macrospin domain avalanche, as shown in \textbf{figure~\ref{fig:No_control_SASI}b}.

\begin{figure*}[t!]

        \begin{subfigure}{\textbf{(a)}}
        \includegraphics[width=0.95\linewidth]{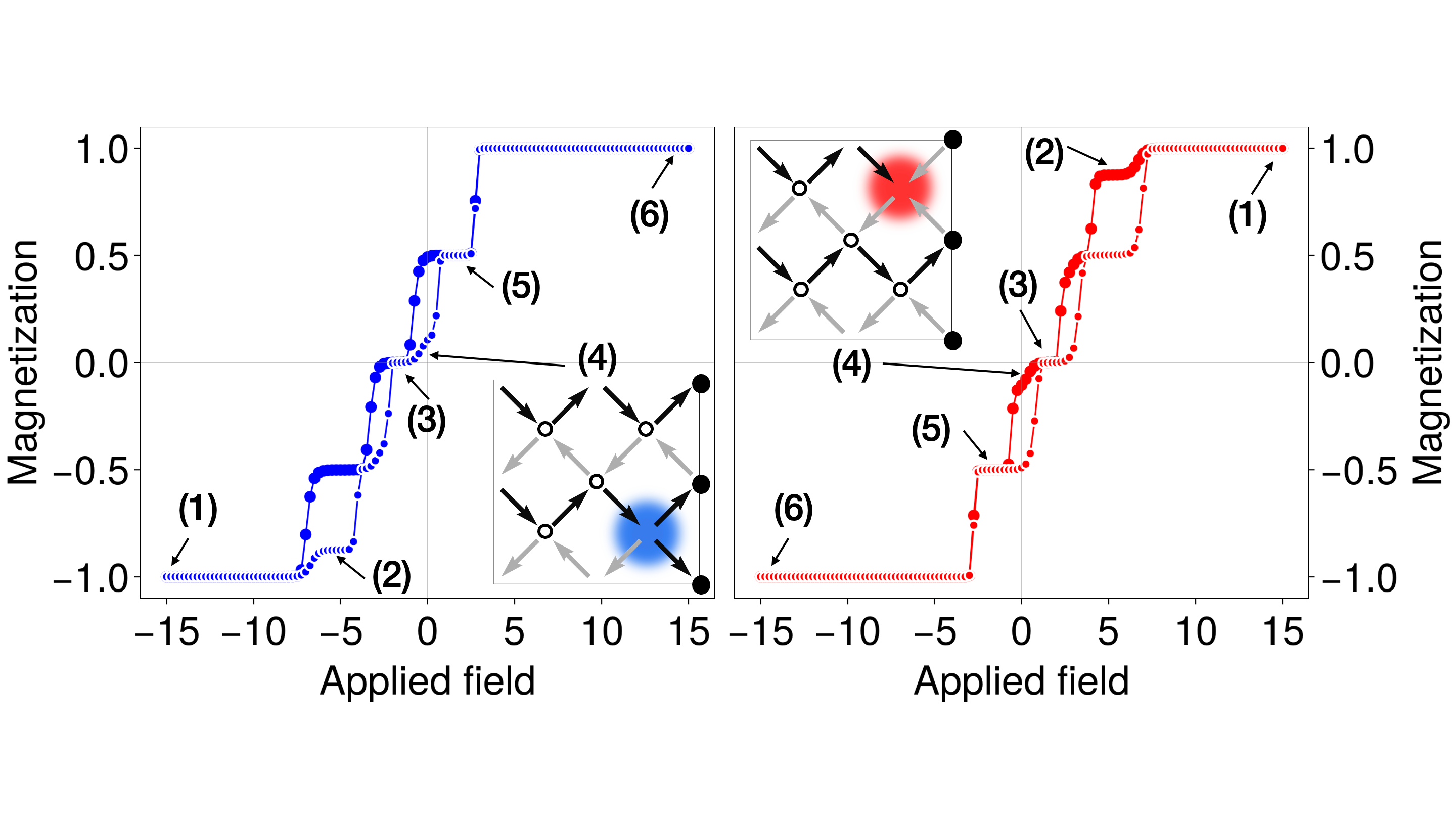}
        \end{subfigure}
        
        \begin{subfigure}{\textbf{(b)}}
        \includegraphics[width=0.95\linewidth]{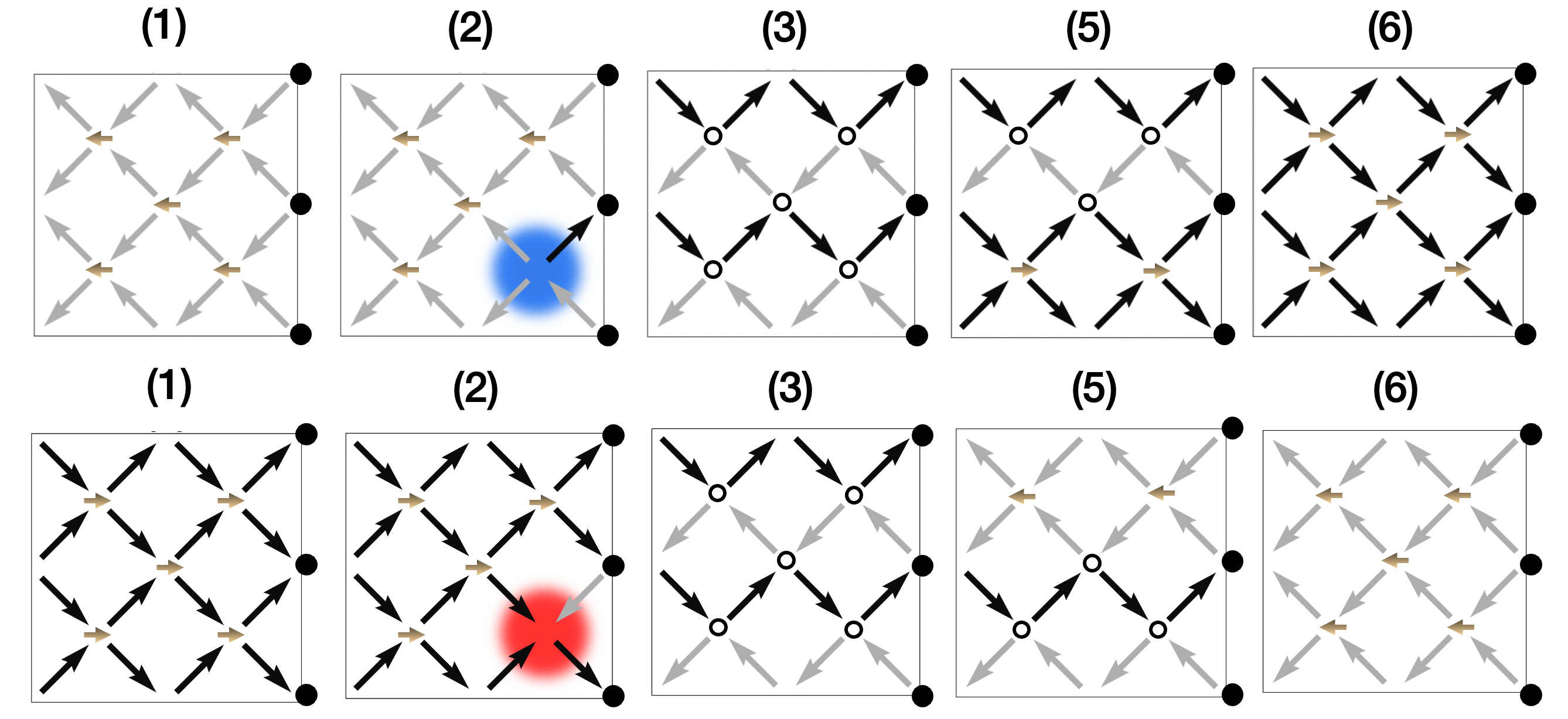}
        \end{subfigure}
        
\caption{\textbf{(a)} Left panel: Average field-dependence of a square ice system with control macrospins (south pole in the lattice plane). Right panel:  Average field-dependence of a square ice system with control macrospins (north pole in the lattice plane). \textbf{(b)} Macrospins and vertex configurations are shown for field regions 1--6 (with the monopole in (2) and (4) (insets in a). Top row: South pole in the lattice plane, inducing $-Q_{m}$ monopoles.  Bottom row: North pole in the lattice plane, inducing $+Q_{m}$ monopoles. The dots ($\bullet$) on the right-most lattice edge are the vertical control macrospins positioned alongside respective lattice edge vertexes. Vertex domain maps (with the macrospin configurations removed) are shown in \textbf{SI4}.\label{fig:Control_SASI}}

\end{figure*}

For the controlled 16-spin ice system, we begin by examining the case where the control macrospins are oriented with their south-facing magnetic poles in the lattice plane—referred to as the ``up'' control configuration. In this setup, the vertical control macrospins placed along the lattice edge generate symmetry-breaking, highly localized magnetic fields that strongly influence domain wall nucleation. These fields affect how Type 3 vertex states emerge and propagate through the lattice. We find that the control macrospins in this configuration impede the nucleation of Type 3 vertices along the edge where they are placed. \textbf{Figure~\ref{fig:Control_SASI}a} (left panel) shows the resulting field-dependent magnetization and corresponding vertex evolution for 16-spin ice system with vertical controls. The individual replicas contributing to the average response are shown in \textbf{SI2}. Reorienting the control macrospins so that the north poles face into the plane (``down'' control) produces a field-dependent magnetization response that mirrors the case shown in Fig.~\ref{fig:Control_SASI}a (left panel).  The outcome of this control reversal is illustrated in the right panel of Fig.~\ref{fig:Control_SASI}a. The introduction of the controls leads to a significantly different field-dependent magnetization response. 

The presence of localized fields from the control macrospins alters both the metastable states the system can occupy and the field values at which magnetization transitions between these states occur. These transitions indicate the reconfiguration of vertex domains and are accompanied by avalanches of macrospin flips and the emergence of Type 3 monopole currents. The mechanism influencing energy barrier for macrospin reversal involves the respective magnetic pole orientations between lattice macrospins at the edge, their neighbouring control macrospins, and the number of macrospins surrounding the edge macrospins.  Also, dipolar fields from the vertical control macrospins will impact the field necessary for a macrospin reversal.

\begin{figure*}[t!]

        \includegraphics[width=0.95\linewidth]{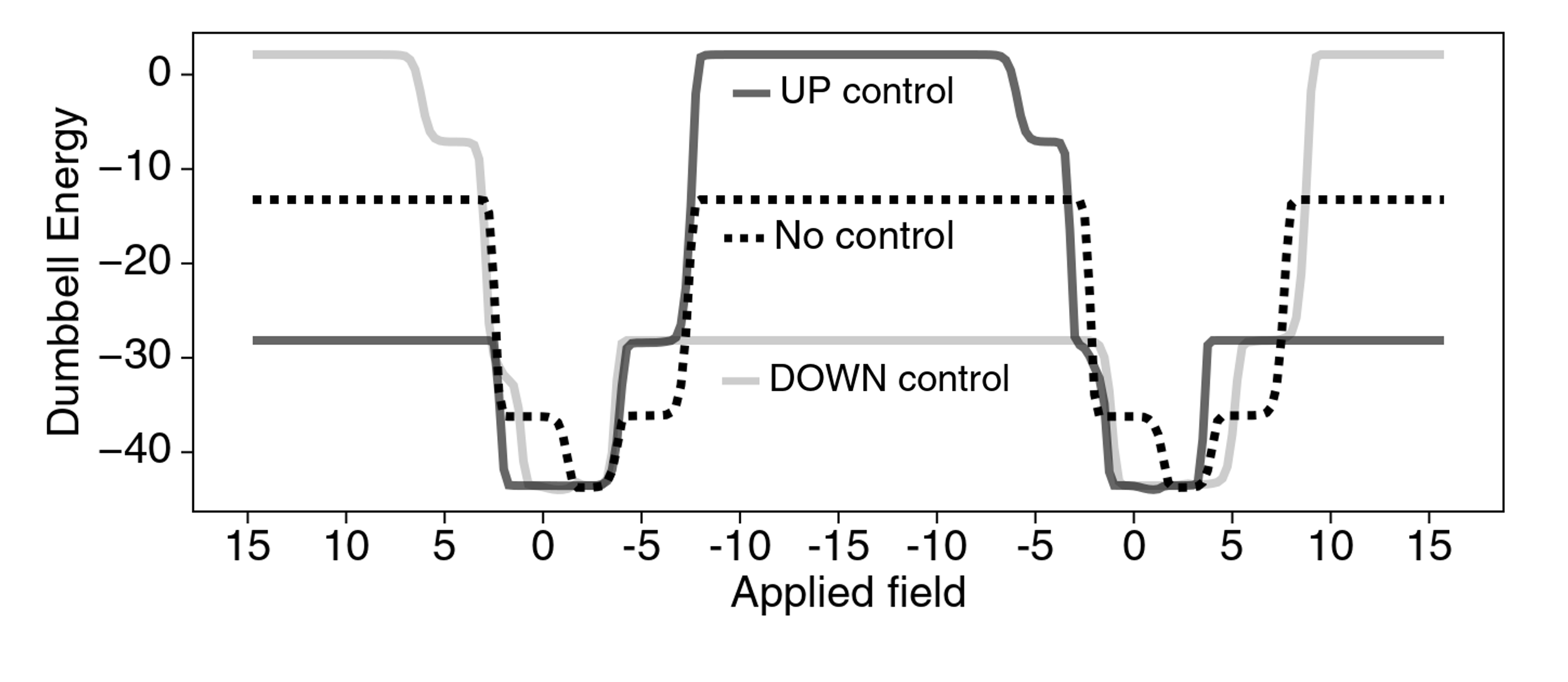}

\caption{ The field dependence of the system's \emph{dumbbell energies}, calculated using the first term in equation 1, is shown averaged over 1000 replicas. The solid lines represent the cases with up and down control, while the dotted line corresponds to the case without control.\label{fig:energy_gradient}}
 
\end{figure*}

To understand how these effects shape the system’s hysteresis response and direct and regulate monopole currents through the lattice at transitions, we examine how the control macrospins modify the magnetostatic energy landscape. The most stable configuration occurs when the number of uncompensated edge macrospins is minimized, allowing for more north–south pole pairings along the lattice perimeter. As shown by the dotted line in Fig.~\ref{fig:energy_gradient}, the field-dependent magnetostatic energy is symmetric with respect to field polarity when no controls are present. However, introducing control elements with south poles in the plane (``up'' control) lowers the energy in high positive fields and increases it in the negative fields, as indicated by the darker solid line in Fig.~\ref{fig:energy_gradient}. This asymmetry arises from the relative alignment of the lattice and control macrospins. In strong positive fields, the control macrospins pin neighbouring edge spins, lowering the energy barrier for reversal along the opposite, non-controlled edge and enabling the nucleation of Type 3 vertices there. As the field decreases, an avalanche of vertex domains occurs, rapidly reversing large regions of the lattice. This collective switching drives the system toward negative saturation magnetization. Under large negative fields, macrospins near the control elements become energetically unfavourable due to the accumulation of positive magnetic charges along that edge. This effect, evident from the relative magnetization directions shown in Fig.~\ref{fig:Control_SASI}b, again reduces the reversal barrier for those spins. As a result, during the return to positive saturation, Type 3 monopoles emerge preferentially at the controlled edge, as seen in configurations (2) and (4) of Fig.~\ref{fig:Control_SASI}. At zero field, all system configurations relax into the same low-energy ground state composed entirely of Type 1 vertices, highlighting the potential for controlling monopole current flow using vertical macrospin orientation.

In addition to monopole flow control, the field-dependent response of a macrospin-controlled artificial spin ice would permit discrete information ``states'' to be stored at each magnetization plateau present in Fig.~\ref{fig:Control_SASI}a. The magnitude of the localized control fields will determine the field at which vertex domains change.  However, suitable local field strengths require a delicate balance between the applied field, the macrospin localized fields, and the dipole inter-element interactions. The localized fields should not be so large that they overwhelm other contributions. Possible implementations include local fields from a dynamically repositionable Magnetic Force Microscopy tip, similar to Wang et al.'s 2016 rewritable magnetic charge ice\cite{Wang.2016}. Also,  permanent localized fields via an additional layer of vertical elements above or below the lattice\cite{Mengotti_2009} with an appropriate materials choice for the control macrospins, e.g. a high-coercivity material would ensure that the macrospin controls maintain their polarity configuration with respect to the lattice. 

Understanding how we can control magnetization dynamics, controlling the flow of the monopole charges,  and access information states in artificial spin ice systems offers potential benefits for various applications. One possibility is making spin ice-based physical reservoir computing and neuromorphic systems\cite{xiang_antiferromagnetic_2024}. By studying multilayered systems with localized controlling fields, we can explore the creation of neural networks and fine-tune accessible states.  Our vertical control strategy also has potential for applications in magnetic memory devices, spintronic devices, and magnetic sensors, where controlled manipulation of magnetic states and charge flow is desirable\cite{heyderman_artificial_2013, jungwirth_antiferromagnetic_2016, stamps_active_2025}. 

Perhaps most interesting, we find a Type 3 vertex monopole current that can be regulated with vertical elements.  All replicas exhibited monopole vertexes via domain wall nucleation sites in the lattice where entire edge vertexes were affected by control macrospins. This result is analogous to what Leon et al.\cite{LEON201659} explored, where the possibility of altering monopole current direction in a square artificial spin ice by heating an edge was presented. These heat-triggered monopole excitations would nucleate at warmed edges, where the monopoles propagate to the opposite edge via the thermal gradient.  In our scheme, the observed monopole current propagates under the influence of the applied magnetic field and is modified by the energy gradient caused by localized fields altering the energy landscape within the system. It is worth noting that thermal effects may influence the impact of macrospin control fields as the critical temperature is approached. Previous investigations of Type 3 vertex monopole current in square spin ice showed that on average the monopole charges travelled in the applied field direction\cite{morley_thermally_2019}. Interestingly, while the square ice with no control macrospins behaves this way, the controlled system results show that we can force the monopole current to flow against the applied field direction\footnotemark{}.\footnotetext{This reversal of monopole current direction occurs only when all vertexes along the controlled edge are occupied by control macrospins. If even a few vertexes on the controlled edge are left open, monopole nucleation can still occur there, leading to current propagation with the applied field direction.} This is because the control macrospins pin edge macrospins, which prevents monopole charges from nucleating along one of the edges. That forces the monopole current to start at the opposite edge and propagate in the direction opposite the applied field.  Additionally, the results in Fig.~\ref{fig:Control_SASI} indicate that south-facing controls favour $-Q_{m}$ monopoles while north-facing controls favour $+Q_{m}$. Thus, our results demonstrate control over both the direction and polarity of monopole currents in artificial spin ice systems, offering a foundation for directional magnetic charge transport in functional device architectures. 

An overall monopole current control is generated that is analogous to how a traditional diode behaves.  The localized fields generated by control macrospins act as a barrier or guide for monopole currents, allowing them to propagate preferentially in one direction. This directionality arises from the polarity of the macrospins and their placement on the lattice. A monopole transistor-like device could also be envisioned, where dynamic switching of monopole current direction or polarity could be achieved by reorienting and/or repositioning the control macrospins. The ability to control monopole currents with localized fields provides an alternative mechanism for encoding, storing, and processing information using magnetic charges instead of traditional electronic currents. The dynamic switching capability has the potential for application to logic operations and programmable spintronic circuits\cite{schiffer_artificial_2021, berchialla_focus_2024, jensen_clocked_2024, marrows_neuromorphic_2024}.

\section{Methods}

\textbf{Monte Carlo Simulation Framework}
Our Monte Carlo simulations used \emph{ReplicaSim}, a custom-built graphics processing unit (GPU) accelerated framework for artificial spin ice systems.  ReplicaSim was written in Julia (v1.11.3) leveraging CUDA.jl and low-level CUDA kernel programming\cite{besard2018juliagpu} for efficient parallel computation. This approach enabled rapid data collection across 1000 replicas, significantly reducing computation time and improving statistics.

\textbf{System Configuration and Parameters}
We modelled the artificial spin ice system as a square lattice of interacting single-domain nanomagnets. Vertical control elements with the same geometric and magnetic properties were added along the edge of the lattice to provide out-of-plane local field control. Each nanomagnet is treated as a dumbbell with a positive and negative magnetic charge at its ends. A magnetic moment vector $\textbf{m}$ is assigned to each dumbbell to represent its macrospin orientation. 
The key simulation parameters were set as follows:
\begin{itemize}
\item [$-$] \textbf{Lattice size per replica}: 16 spins. 
\item [$-$] \textbf{Replica count}: 1000 (simulated in parallel on a single GPU).
\item [$-$] \textbf{Monte Carlo steps per field}: 1000 steps at each of 303 field values covering a full hysteresis cycle.
\item [$-$] \textbf{Sampling note}: Only spins indexed 1–16 (lattice macrospins) were sampled for reversal. Control macrospins (indices 17–19) were fixed and excluded from sampling but included in the dipolar interaction calculations.
\item [$-$] \textbf{Temperature}: $T$ = 0.3 (in reduced energy units).
\item[$-$] \textbf{Interaction model}: Long-range dipolar interactions via a modified dumbbell model with precomputed neighbor lists.
\item [$-$] \textbf{Neighbour cutoff radius}:  $R_c$ = 6 lattice units, including up to 112 neighbours per spin. We determined this cutoff by comparing the total energy of a system containing $N$ = 100 spins with and without various cutoffs and choosing the smallest radius that yielded <1\% difference in energy from the full calculation (see \textbf{SI5}). Note that the optimal cutoff radius may differ for other lattice geometries or nanoelement arrangements. For the small 16-spin system studied here, all spins are effectively included, capturing finite-size effects.
\item [$-$] \textbf{Field-sweep protocol (hysteresis loop)}: Replicas were initialized randomly, saturated in positive field, and swept from positive to negative saturation and back in steps of 0.25. At each step, 1000 Monte Carlo steps were run, and magnetization, energy, and configurations were recorded. The applied field angle with respect to the $x$-axis was $\phi=0.1^{\circ}$, which sufficiently removed the degeneracy of states.
\end{itemize}

Before obtaining the data, we verified that 1000 Monte Carlo steps per each applied field value in the hysteresis simulation was an appropriate number of steps to reach equilibrium. Magnetization values stabilize after approximately 500 steps, as shown in \textbf{SI6}.

\begin{suppinfo}

Field-dependent magnetization data for 1000 replicas of square artificial spin ice with and without control macrospins; comparison of magnetization for south- and north-facing control configurations; cutoff radius optimization; and Monte Carlo equilibration analysis.
	
\end{suppinfo}

\section{Notes}
The authors declare no competing financial interest.\\
The \emph{ReplicaSim} code used for GPU-accelerated Monte Carlo simulations and the supporting documentation are publicly available on GitHub:  https://github.com/ju-pixel/ReplicaSim.\\

\begin{acknowledgement}

This work was supported by the University of Manitoba, the Natural Sciences and Engineering Research Council of Canada (NSERC RGPIN 05011-18 and RGPIN-2024-04882), and the Canadian Foundation for Innovation (CFI) John R. Evans Leaders Fund.

\end{acknowledgement}

\bibliography{SASI_MvsH.bib}

\appendix
\newpage
\section{Supporting Information}

\begin{figure}

\includegraphics[width=0.95\linewidth]{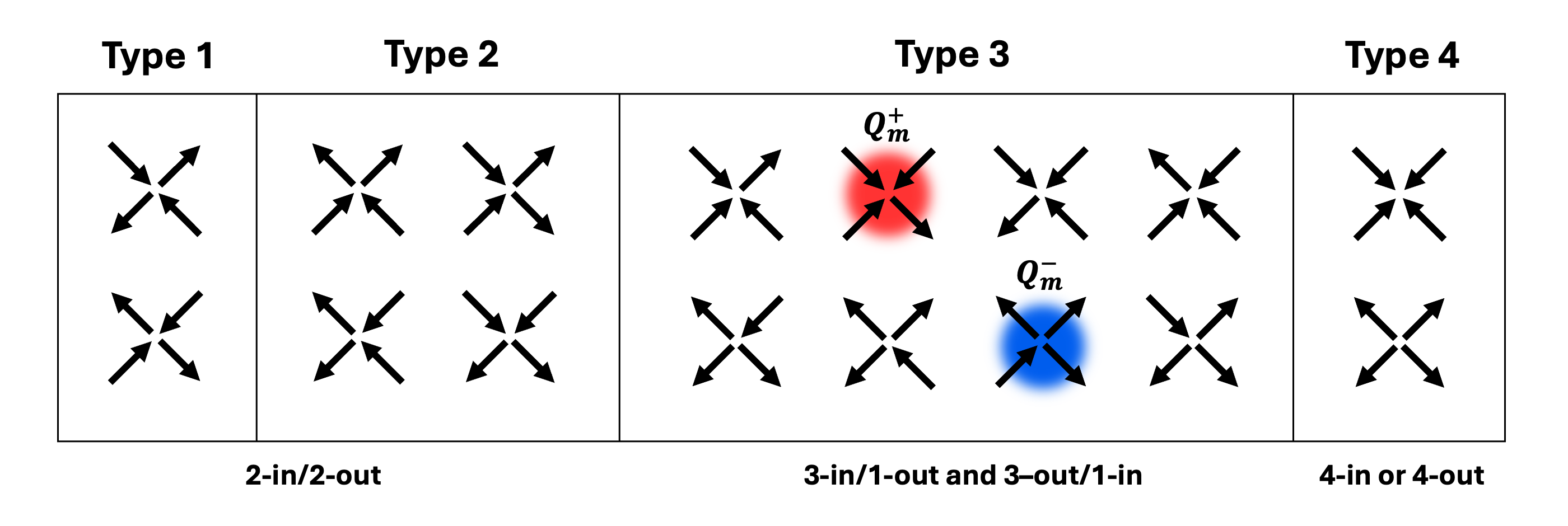}
\caption{Macrospin configurations of a square artificial spin ice vertex grouped into the four subtypes: Type 1, Type 2, Type 3 (also known as monopole vertex configuration), and Type 4.}

\end{figure}

\newpage

\begin{figure}

\includegraphics[width=0.95\linewidth]{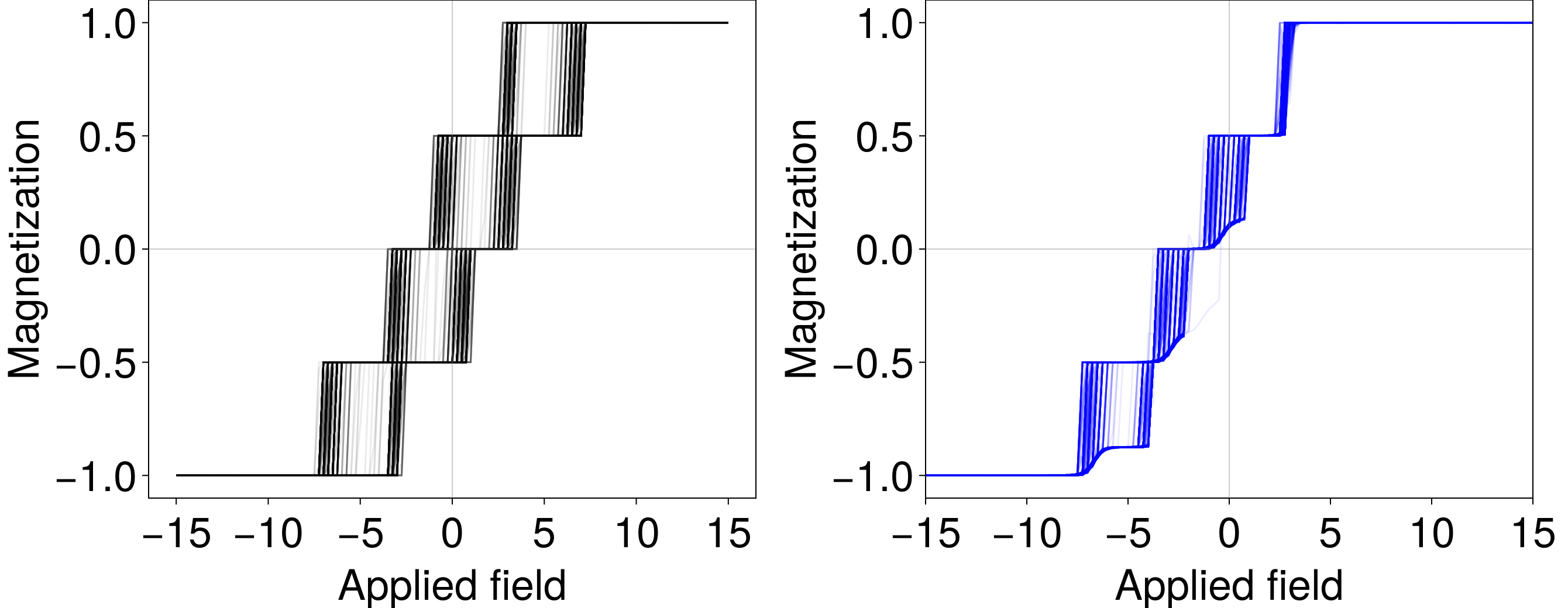}

\caption{Left: Hysteresis loops of 1000 replicas; 16-spin ice. Right: Hysteresis loops of 1000 replicas; 16-spin ice with control macrospins having a south pole in lattice plane (``up'' control).}

\end{figure}

\begin{figure}

\includegraphics[width=0.95\linewidth]{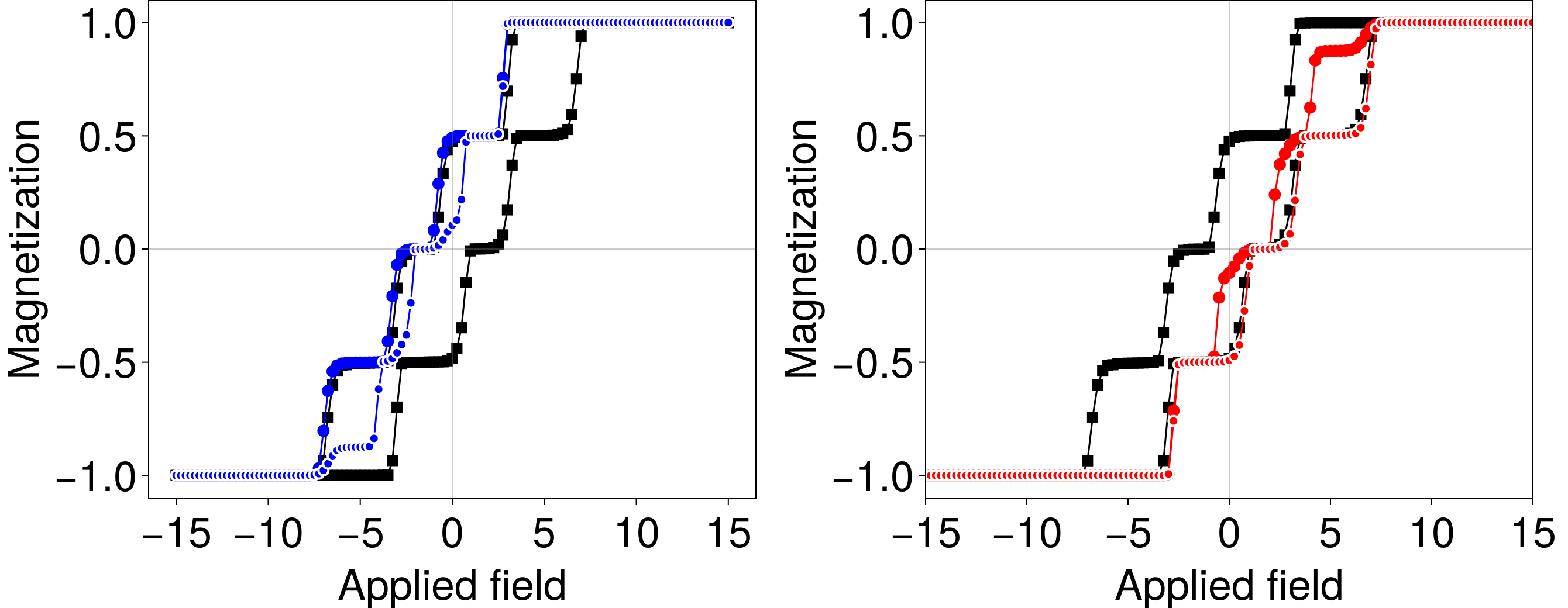}

\caption{Average hysteresis loops for square ice, black $\Box$.  Left panel is when the system has control macrospin south poles in the lattice plane (``up'' control), blue $\bigcirc$; and the right panel is when the control macrospin north poles are in the lattice plane (``down'' control), red $\bigcirc$.}

\end{figure}

\begin{figure}

\includegraphics[width=0.95\linewidth]{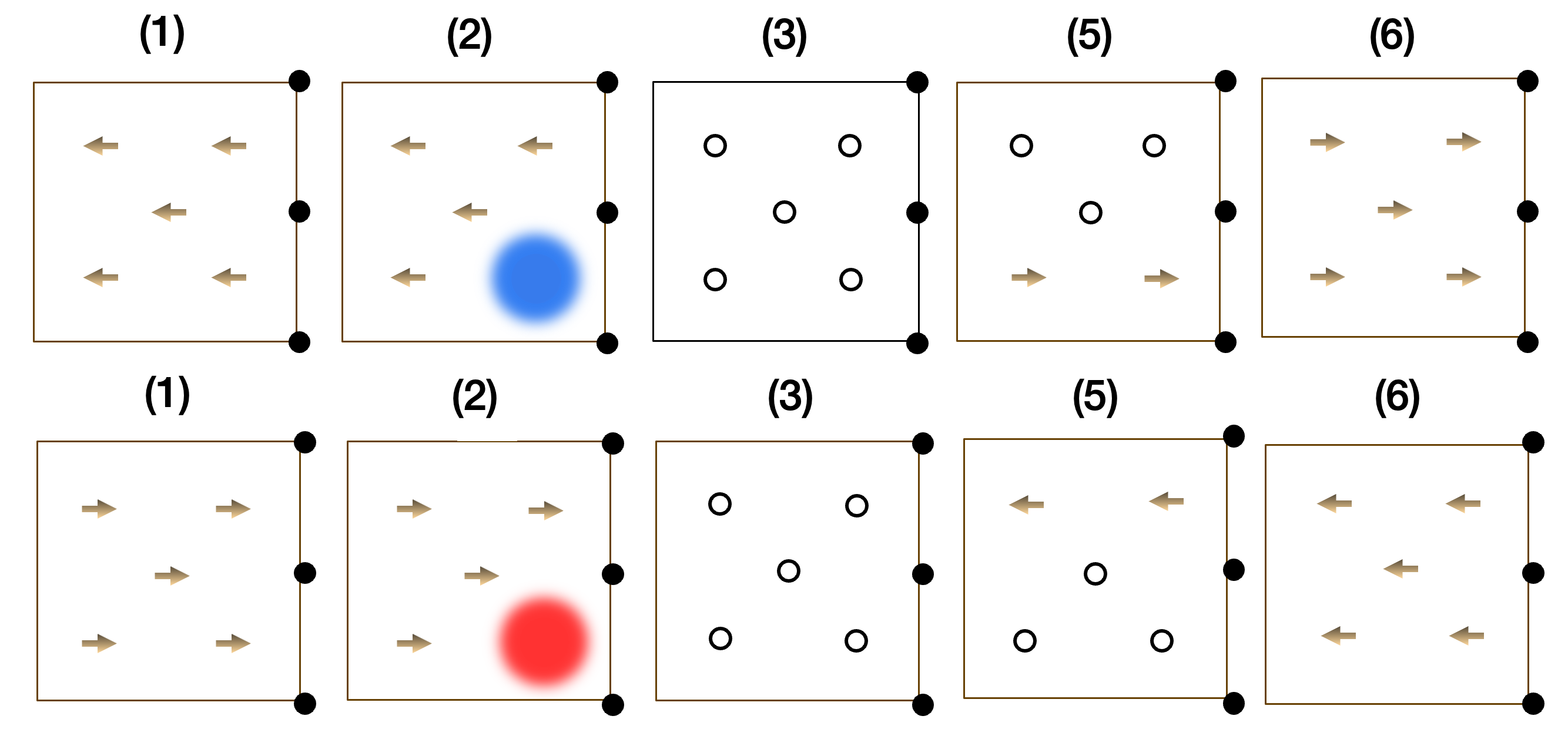}

\caption{Vertex domains at selected field values in the hysteresis loop for the controlled system, shown for a single replica out of 1000. The top row corresponds to the case where the control macrospin south poles are in the lattice plane (``up'' control), while the bottom row corresponds to the control macrospin north poles in the lattice plane (``down'' control).  Macrospin configurations were removed for clarity. Type 2 vertices are indicated by brown arrows ($\leftarrow$ and $\rightarrow$), and Type 1 vertices are marked by black circles ( $\bigcirc$.) The large blue and red fuzzy circles in snapshot (2) represent a Type 3 vertex/monopole charge. Snapshot (3) shows a perfect ground state consisting of only Type 1 vertexes for both control cases.}

\end{figure}

\begin{figure}

\includegraphics[width=0.75\linewidth]{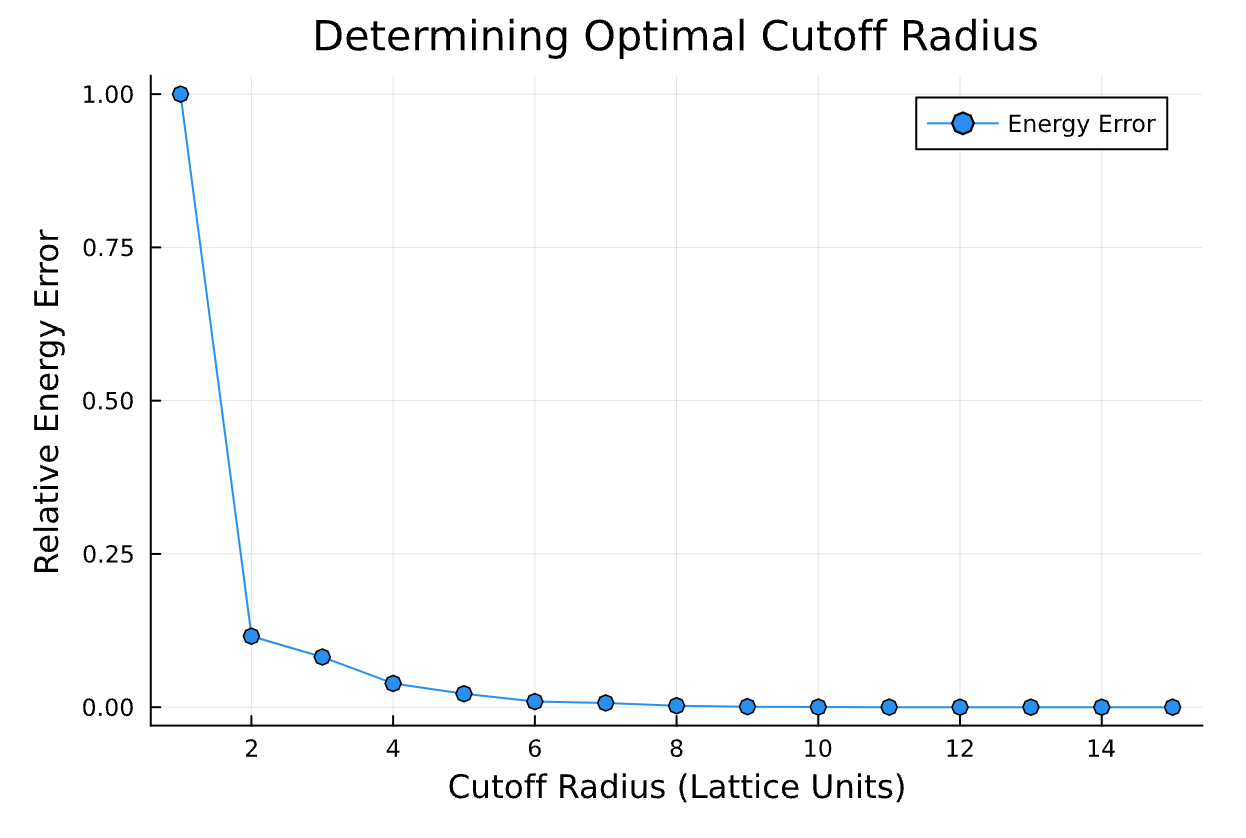}

\caption{Determining the optimal cutoff radius for dipolar interactions in artificial spin ice simulations. The plot shows the relative energy error as a function of the cutoff radius (in lattice units). The energy error is computed as: 
$ \text{Energy Error} = \frac{|E_{cutoff} - E_{full}|}{|E_{full}|}$
where $E_{cutoff}$is the total system energy calculated with a finite cutoff radius, and $E_{full}$ is the reference energy computed with an effectively infinite interaction range. The error decreases rapidly with increasing cutoff, stabilizing around $R_c \approx $ 6-8  lattice units, beyond which additional terms contribute negligibly to the total energy. This analysis helps balance computational efficiency and accuracy in Monte Carlo simulations of artificial spin ice.}

\end{figure}

\begin{figure}

\includegraphics[width=0.70\linewidth]{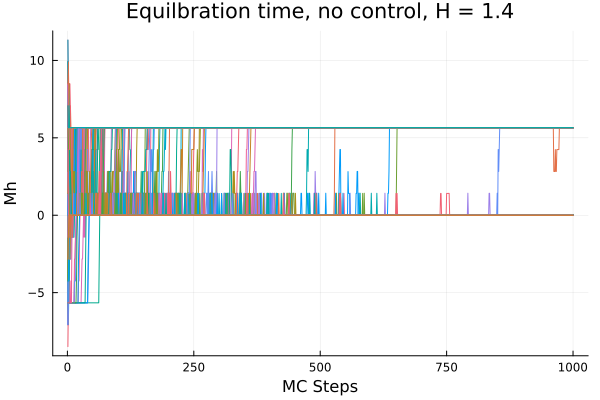}
\vfill
\vspace*{25pt}
\includegraphics[width=0.70\linewidth]{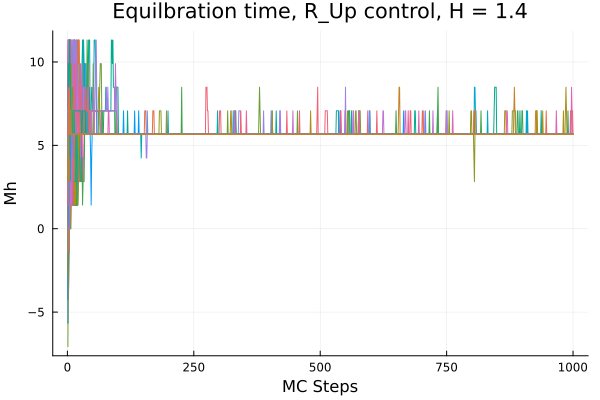}

\caption{Equilibration dynamics of the system with and without control elements at $T$ = 0.3 and $H$ = 1.4, a field near a critical value for this system.
\textbf{Top panel}: The magnetization $M_{h}$, projected onto the field direction, is plotted as a function of Monte Carlo steps in a system without control elements. Each coloured line represents one of the 1000 independent replicas. The magnetization fluctuates significantly before stabilizing after approximately 500 Monte Carlo steps, indicating each replica settling into one of the two preferred metastable states.
\textbf{Bottom panel}: The same system, now with control elements. The presence of controls significantly alters the equilibration behaviour. Some replicas rapidly stabilize, while others remain trapped in a preferred meta-stable state with occasional fluctuations persisting, compared to the noncontrolled case. This suggests that the localized field from the control elements influences the magnetization relaxation dynamics and may introduce a preferred energy configuration that constrains the system’s evolution.}

\end{figure}

\end{document}